# Diffusion Entropy Approach to Dynamical Characteristics of a Hodgkin-Huxley Neuron

Huijie Yang[☆], Fangcui Zhao, Zhongnan Li, Wei Zhang

*School of Physics, Nankai University, Tianjin 300071, China*
[☆] *Corresponding author, E-mail: huijieyangn@eyou.com*

**Abstract.** By means of the concept DE we analyze the responses of a HH neuron to two types of spike-train inputs. Two characteristic quantities can be extracted, which can reflect partially the dynamical process of a HH neuron.

**PACS number(s):** 87.18.Sn, 84.35. +I, 05.45.Tp

## I. INTRODUCTION

All types of information received by sensory system are encoded by nerve cells into sequences of spikes before they are transmitted to the brain. The brain neurons use these resulting sequences as main instrument for intercells connection. The information is modulated into the series of the interspike intervals of the action potential train (ISIs). Extracting dynamical characteristics from ISIs is an essential role to understand the mechanism of a neuron.

Very recently, diffusion entropy (DE) is designed to analyze time series [1-4]. It can reveal the dynamical characteristics of a dynamical system from complex to equilibrium. DE is much more effective in detecting scaling invariance rather than variance-based methods. In this paper we try to analyze the response ISIs of a Hodgkin-Huxley (HH) [5] neuron to two types of spike-train inputs.

## II. DE ANALYSIS [1-4]

Denoting the output ISIs as, $\{T_o(k) | k = 1, 2, ..., N\}$. Connecting the starting and the end of this time series, we can obtain a set of delay register vectors as,

$$\{T_0(1), T_0(2), \cdots T_0(n)\}$$
$$\{T_0(2), T_0(3), \cdots T_0(n+1)\}$$

$$\vdots$$

$$\{T_0(N), T_0(1), \cdots T_0(n-1)\} \tag{1}$$

Considering each vector as a trajectory of a particle in duration of $n$ time units, all the above vectors can be regarded as a diffusion process for a system with $N$ particles. Accordingly, for each time denoted with $n$ we can reckon the distribution of the displacements of all the particles as the state of the system at time $n$. Dividing the possible range of displacements into $M_0$ bins, DE approach defines diffusion entropy as,

$$S(n) = -\sum_{m=1}^{M_0} \frac{K_m(n)}{N} \ln\left(\frac{K_m(n)}{N}\right). \tag{2}$$

Where $K_m(n) | m = 1, 2, \ldots M_0$ is the number of particles whose displacements fall in the $m$'th bin at time $n$.

To get a suitable $M_0$ the size of a bin $\varepsilon$ can be a fraction of the variance of the ISIs, which reads,

$$\varepsilon = \sqrt{\frac{\sum_{m=1}^{N} T_o(m)^2}{N}}.$$

According to a tenet of the Science of Complexity [6,7], complexity is related to the concept of diffusion scaling. Diffusion scaling can be defined by,

$$p(m,n) = \frac{K_m(n)}{N} = \frac{1}{n^\delta} F\left(\frac{m}{n^\delta}\right) \Big| m = 1, 2, \ldots M_0. \tag{3}$$

Assuming the probability distribution function (PDF) of a diffusion process fulfills this scaling property, it can be proved via some trivial computation that the diffusion entropy obeys a relation as follows,

$$S(n) = A + \delta \cdot \ln n, \tag{4}$$

Where $A$ is a constant depending on the function form of PDF.

Hence, DE can give a reliable value of the scaling exponent $\delta$ rather than the second moment of the diffusion process, $<x^2> \propto n^{2H}$. DE is sensitive to complexity of a dynamical system. Complex systems are expected to generate a departure from the condition of ordinary diffusion, where $\delta = 0.5$ and the function $F\left(\frac{m}{n^\delta}\right)$ in the PDF is a Gaussian function of $\frac{m}{n^{1/2}}$. DE can be used as a reliable tool to obtain the important measure of complexity of a dynamical system, $\delta$.

Because of the periodic condition the displacements at time $n$ can be written as,

$$D_s(n) = \sum_{i=s}^{s+n-1} T_0(i) \Big| s = 1, 2, 3, \ldots N. \tag{5}$$

On the other hand, the displacements at time $N - n$ are,

$$D_s(N-n) = \sum_{i=s}^{s+N-n-1} T_0(i)) \Big| s = 1,2,3,...,N$$

$$= \sum_{i=s}^{s+N-1} T_0(i) - \sum_{i=s+N-n}^{s+N-1} T_0(i) \Big| s = 1,2,3,...,N$$

$$= T_c - \sum_{i=s+N-n}^{s+N-1} T_0(i) \Big| s = 1,2,3,...,N \qquad (6)$$

$$= -D_{s+N-n}(n) + T_c \Big| s = 1,2,3,...,N$$

Where $T_c$ is the summary of all the elements, and can be regarded as a constant. Hence the PDF at time $n$ is identical with that at time $N-n$, the DE results are symmetric with respect to the time point $n = N/2$.

## III. RESULTS

The response output of a HH neuron to time-independent ISIs is presented in Fig. (1a). And the corresponding DE result for output ISIs is shown in Fig. (1b). For this kind of stimuli, we are mainly interested in the conditions with larger values of $T_i$, at which the output ISIs may be much more informative [8]. $T_i$ is assigned $30m\sec$ in calculations.

The response output of a HH neuron to time-dependent ISIs modulated by sinusoidal signal is also calculated. The input ISIs obeys $T_i = c + d\sin(2\pi t/T_p)$. In Fig. (2), $c, d$ and $T_p$ are chosen to be $20m\sec, 10m\sec, 100m\sec$, while in Fig. (3) $20m\sec, 10m\sec, 50m\sec$, and in Fig.(4) $10m\sec, 5m\sec, 100m\sec$.

Though the dynamical process for a long time is simulated, we are mainly interested in the time range of several seconds. Because the average ISI is about $10m\sec$, the corresponding scale we are interested in the DE results is $n \sim 10^2$.

In all the cases the DE of the diffusion process generated by the time series shows a damped oscillatory behavior with a partial regression to the initial condition. The minima of this regression process, represented in a linear-log plot, lie on a straight line, whose slope corresponds to an anomalous scaling. As suggested in Ref. [2], the slope of the bottom line is the scaling index that would correspond to the neuron equilibrium. However, the neuron response lives in a condition, intermediate between the dynamic and thermodynamic state, as indicated by the fact the entropy maximum values undergo a slow process of regression, which is another inverse power law regression, to the condition of equilibrium indicated by the bottom line. This slow process of regression is interrupted by saturation phenomena at large time, probably caused the fact that the time series is finite.

Accompanying with this anomalous scaling $\delta$, we can find that a DE result recurrences with almost an exact period $T_{HH}$. The values of these two measures for the conditions of time-independent ISIs and ISIs modulated by sinusoidal signal are,

$$(T_{HH}, \delta)_{T_i=30m\sec} = (38, 0.56),$$

$$(T_{HH}, \delta)_{T_i=[20+10\sin(2\pi/100)]} = (58, 0.83),$$

$$(T_{HH}, \delta)_{T_i=[20+10\sin(2\pi/50)]} = (38, 0.82),$$

$$(T_{HH}, \delta)_{T_i=[10+5\sin(2\pi/100)]} = (42, 0.90) \qquad (7)$$

We can notice that the periodicity emerging from the use of the DE of the diffusion process is not trivially related to the periodicity of the stimulus.

DE approach can detect two basic characteristics of the output ISIs of a HH neuron under stimuli, i.e., the recurrence period $T_{HH}$ and the complexity index $\delta$. The dynamics of a HH neuron is a complexity state, intermediate between dynamics and thermodynamics.

## ACKNOWLEDGMENTS

This work was supported by the National Science Foundation of China under Grant No.60274051/F0303 and No.10175036. One of the authors (F.C. Zhao) would also thank the Post-doctor Fund of Nankai University.

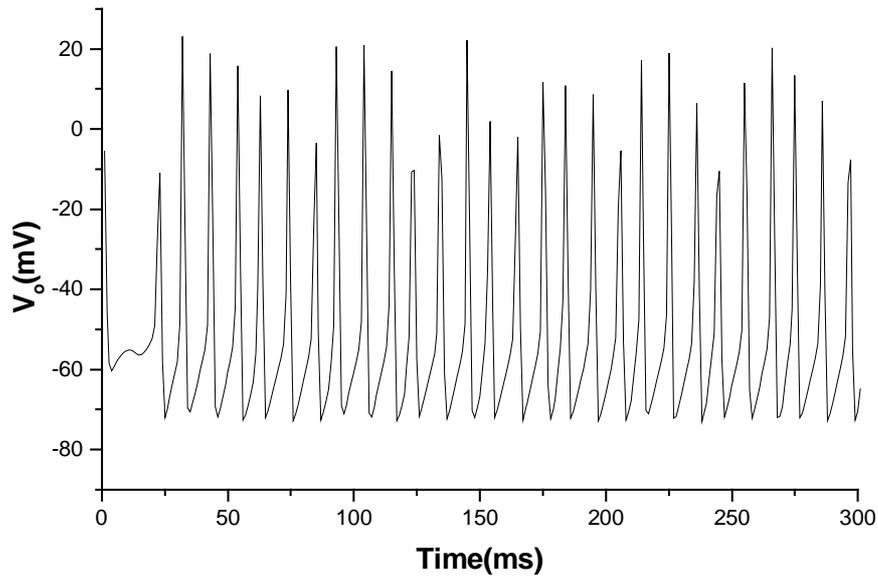

**Fig.(1a)** Output series for a HH neuron stimulated with a equal-interval burst signal. $I_s=25\mu A/cm^2$, $T_i=30$msec. Result In duration $0msec<t<300msec$ is presented.

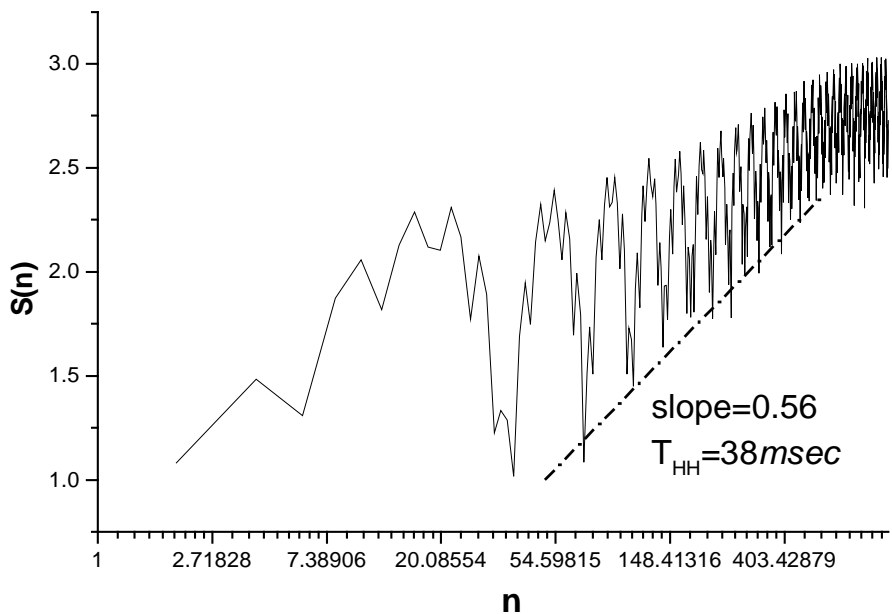

slope=0.56
$T_{HH}=38msec$

**Fig.(1b)** DEA result for output ISIs. The stimulating series is a equal-interval burst signal. $I_s=25msec$, $T_i=30msec$.

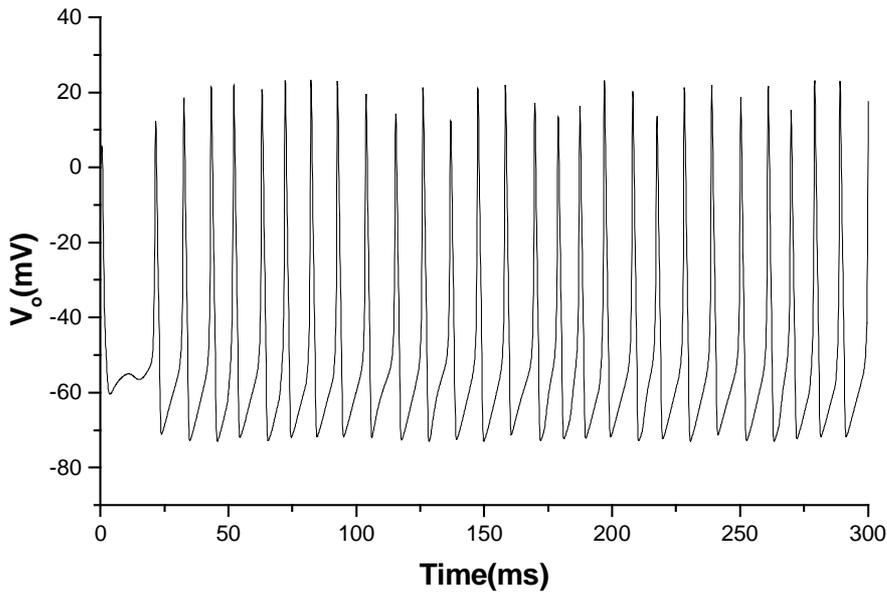

**Fig.(2a)** Output series for a HH neuron stimulated with a sin-modulated interval burst signal. $I_s=25\mu A/cm^2$. The input ISIs obeys $c+d\sin(2\pi t/T_p)$. The condition $c=20msec$, $d=10msec$, $T_p=100msec$ is simulated. $0msec<t<300msec$ is presented.

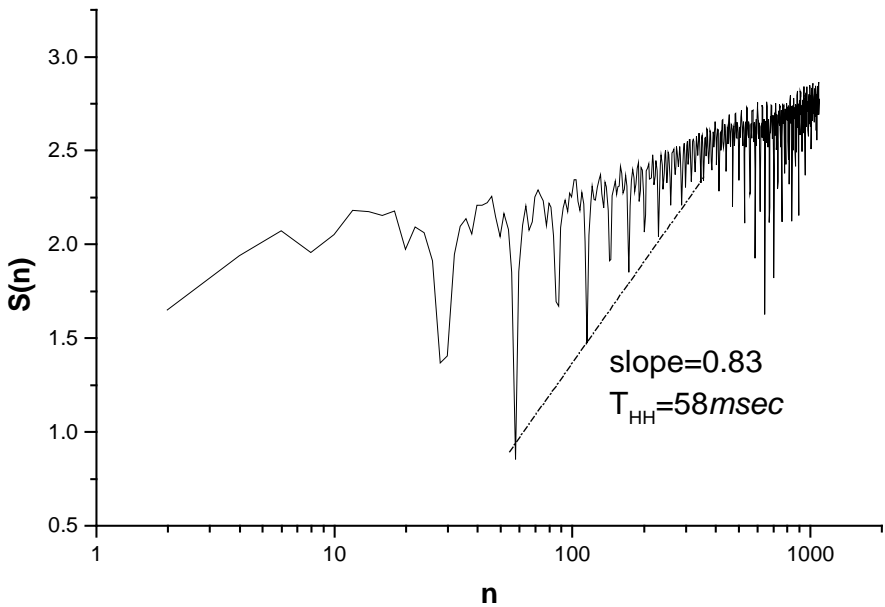

**Fig.(2b)** DEA result for output ISIs. The stimulating series is a sin-modulated interval burst signal. The input signal obeys $c+d\sin(2\pi t/T_p)$. The condition $c=20msec$, $d=10msec$, $T_p=100msec$ is simulated.

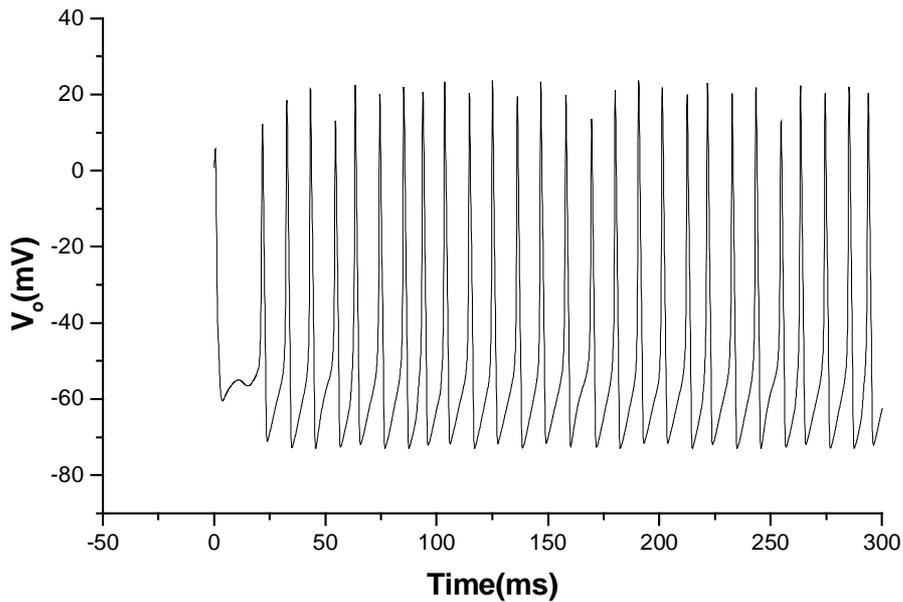

**Fig.(3a)** Output series for a HH neuron stimulated with a sin-modulated interval burst signal. $I_s=25\mu A/cm^2$. The input ISIs obeys $c+d\sin(2\pi t/T_p)$. The condition $c=20msec, d=10msec, T_p=50msec$ is simulated. $0msec<t<300msec$ is presented.

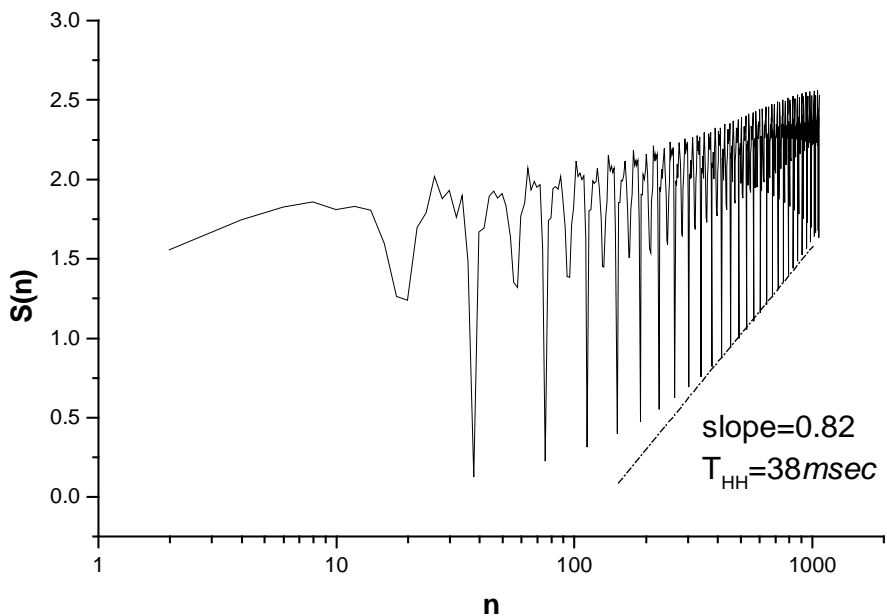

**Fig.(3b)** DEA result for output ISIs. The stimulating series is a sin-modulated interval burst signal. The input signal obeys $c+d\sin(2\pi t/T_p)$. The condition $c=20msec, d=10msec, T_p=50msec$ is simulated.

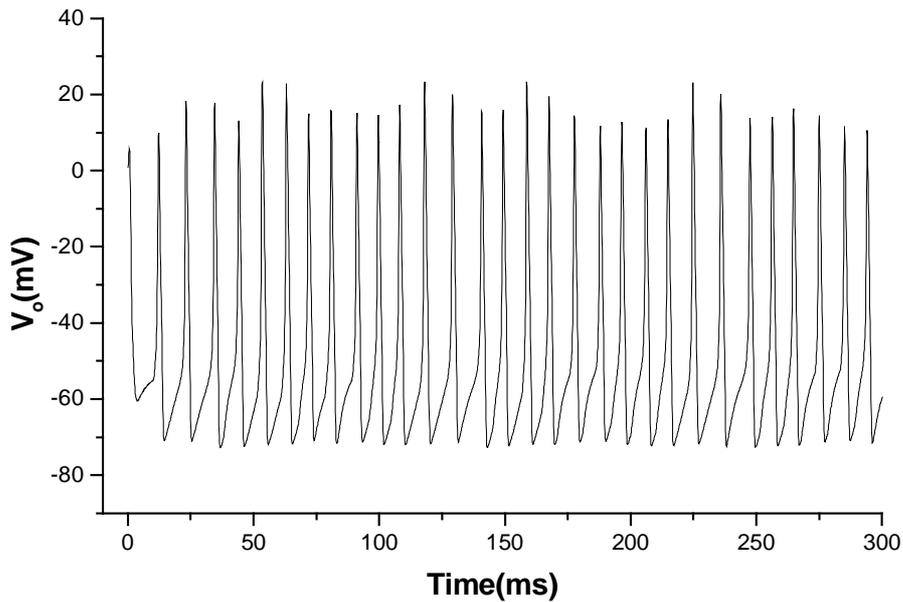

**Fig.(4a)** Output series for a HH neuron stimulated with a sin-modulated interval burst signal. $I_s=25\mu A/cm^2$. The input ISIs obeys $c+d\sin(2\pi t/T_p)$. The condition $c=10msec$, $d=5msec$, $T_p=100msec$ is simulated. $0msec<t<300msec$ is presented.

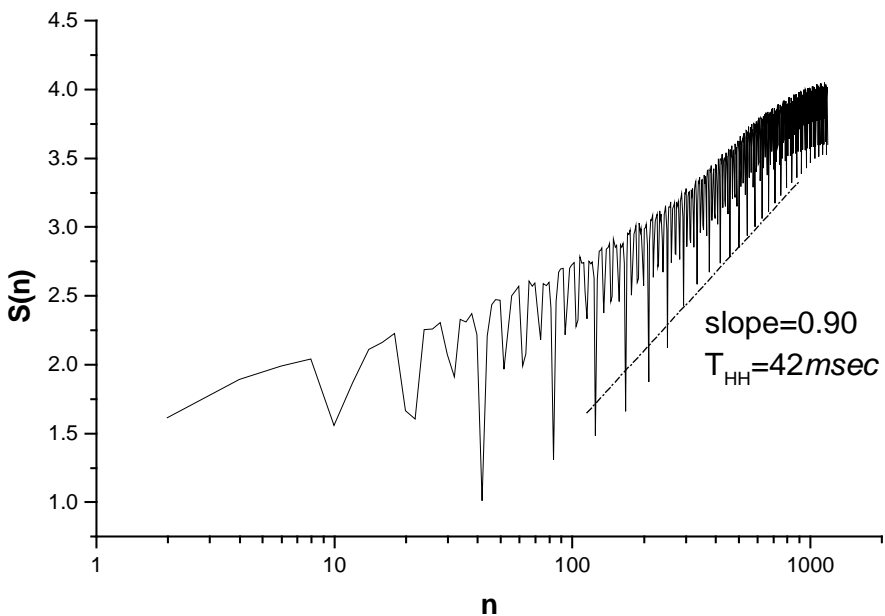

**Fig.(4b)** DEA result for output ISIs. The stimulating series is a sin-modulated interval burst signal. The input signal obeys $c+d\sin(2\pi t/T_p)$. The condition $c=10msec$, $d=5msec$, $T_p=100msec$ is simulated.